# Predicting Friction under Vastly Different Lubrication Scenarios

Yulong Li,[1,2]* Peter Gumbsch,[1,3] Christian Greiner[1,2]†

[1]Institute for Applied Materials (IAM), Karlsruhe Institute of Technology (KIT), Kaiserstrasse 12, 76131, Germany
[2]MicroTribology Center (µTC), Strasse am Forum 5, 76131, Karlsruhe, Germany.
[3]Fraunhofer IWM, Wohlerstrasse 11, 79108, Freiburg, Germany.

*Corresponding author
Email: christian.greiner@kit.edu, yulong.li@kit.edu

**ABSTRACT**. Friction is ubiquitous in daily life, from nanoscale machines to large engineering components. By probing the intricate interplay between system parameters and frictional behavior, scientists seek to unveil the underlying mechanisms that enable prediction and control of friction - an essential step toward carbon neutrality. Yet, reproducing frictional behavior in experiments is notoriously difficult. Here, we show that this challenge stems from the extreme sensitivity of tribological systems to tiny variations, e.g. in surface topography, typically presumed well-controlled. Even after meticulous surface preparation to semiconductor-industry standards and curtailing misalignment-induced oscillations, subtle variations remain and interact. In turn, such minute initial differences lead to statistically significant variations in friction and wear, giving rise to system-level chaotic behavior. Yet, by leveraging mid-scale features of surface topography and misalignment-induced oscillations - information often filtered out or overlooked - we established a model that accurately predicts high-friction regions under vastly different lubrication scenarios, with its performance further enhanced by machine learning.

## I. INTRODUCTION.

Researchers estimate that about 23% of mankind's primary energy usage goes into overcoming friction forces and causing material wear [1]. This highlights the critical role of tribology - the study of friction, wear, and lubrication - in efforts to reduce $CO_2$ emissions. Since Leonardo da Vinci first articulated the 'laws' of sliding friction in 1493, scientists have strived to predict friction in order to build on existing results and design tailored low-friction systems - an endeavor with the potential to save a quad of energy ($10^{15}$ BTU) [2], equivalent to $10^{18}$ J. However every scientist studying tribological systems knows how extremely hard, or truth be told, impossible it is to precisely reproduce frictional behavior. This irresistibly calls to Eduard Lorenz's description of chaos: "When the present determines the future but the approximate present does not approximately determine the future." Such complexity of tribological experiments is especially problematic as tribology often lacks attention and remains a small field of study, while being of immense importance. Friction, a fundamental aspect of tribology, presents significant reproducibility challenges [3], making prediction even more difficult: minor deviations in environmental conditions [4], material properties [5], or surface topography [6] can lead to variations of up to 20% between laboratories and up to 13% within the same lab, even when researchers strictly control conditions. Beyond various strategies to enhance reproducibility through improved experimental and statistical design [3,7], uncovering the root causes of irreproducibility may reveal new predictors for friction behavior - an insight that lies at the core of this study's contribution.

## III. RESULTS

The inescapable interplay between frictional behavior and surface topography is present in every tribological system [8]. Surface topography, shaped by various physical effects and often exhibiting self-affine fractal properties, is highly complex and inherently irreproducible in fine detail [9,10]. This characteristic may lead tribological systems to exhibit chaotic behavior, where minute changes in initial conditions - such as the surface topography of the contacting partners - may exert unexpected and significant effects. To probe the above, we performed tribological experiments that were on purpose very fundamental and basic, while reducing potential influencing factors. We chose a classical pin-on-disk setup (in Fig. 1(a)) and one of the most common bearing steels, 100Cr6 aka AISI 5210, for both the pin and the disk side. A flat-ended pin was deliberately selected over a spherical one to create a well-defined, constant contact area with lower nominal contact stress. This design choice minimizes the influence of an evolving pin geometry during test and allows the study to focus on how disk surface topography affects friction behavior. To ensure the broad applicability of our results, we performed tests for a wide variety of tribology conditions at room temperature. These were: With a non-additive base oil, a fully formulated engine oil, an aerospace grease, and an alumina slurry.

*Contact author: yulong.li@kit.edu
†Contact author: christian.greiner@kit.edu

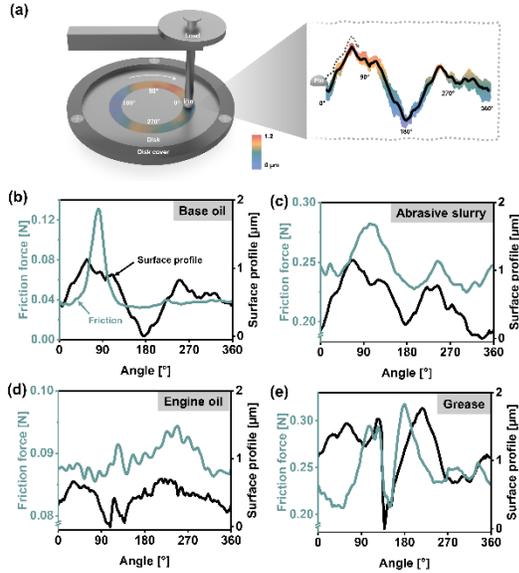

**FIG 1. Comparison of disk surface profile and friction.** (a) Pin-on-disk setup. The disk rotates clockwise with the normal load applied to the fixed pin. The disk surface profile along the contact area with the pin is extracted from chromatic white light profilometry data. (b-e) Comparing disk angle-resolved friction forces with the disk surface profile under different lubrication scenarios - base oil (b), abrasive slurry (c), engine oil (d), and grease (e). The disk angle-resolved friction force was evaluated over a sliding distance of 0-50 m (approximately 379 disk revolutions) by segmenting the entire 132 mm sliding track (360°) into 120 equal sections.

### Surface topography's influence on friction

Given that every surface topography is unique, we aimed to minimize differences between them by keeping the disks as flat as possible, ensuring that the surface profiles (in Fig. 1(a-e) and Fig. S1) height variation of all disks was less than 2 μm along 132 mm sliding track. The three-dimensional topography data, acquired via chromatic white light profilometry, were angularly averaged to yield a two-dimensional profile representation (Fig. 1(a)). An example profile from a disk with a maximum height variation of 1.2 μm over the 132 mm radius is shown in Fig. 1(a and b). Surface profiles were sampled at 120 evenly spaced angular positions along the track, providing a spatial resolution of 1.1 mm. This implies that the analysis captures and focuses on mid-scale surface topography, while suppressing small-scale roughness. In highly precise semiconductor manufacturing, e.g., for the 7 nm node lithography process, a height variation of up to 5 μm is tolerated on a 300 mm wafer [11], demonstrating the precision of our disk preparation in comparison; our disks for all intents and purposes can be considered "flat" as far as tribological experiments are concerned. When plotting the disk angle-resolved friction forces (evaluated over a sliding distance of 0-50 m) with the disk surface profile presented in Fig. 1(b), the peak in the angularly resolved friction force (30°-120°) coincides with the peak in the surface topography (10°-150°). Apart from its height, the surface texture at the peak shows no significant difference from other regions. This indicates that even tiny deviations on the surface (1 μm along a 132 mm sliding track in Fig. 1(b)) can significantly increase friction, by up to 300% in this case. In a similar vein, the high friction section also shows more wear scratches than the low friction section (in Fig. S2). The correlation between surface profile height and friction emerges as a pervasive characteristic across diverse lubrication scenarios - base oil (around 90° in Fig. 1(b)), abrasive slurry (around 100° in Fig. 1(c)), engine oil (around 230° in Fig. 1(d)), and grease (around 120° in Fig. 1(e)) - despite nearly an order of magnitude differences in their absolute friction levels.

*Contact author: yulong.li@kit.edu
†Contact author: christian.greiner@kit.edu

However, comparing friction across different lubrication scenarios poses challenges. As shown in Fig. 2(a), both friction force values and their trends differ significantly under abrasive wear and base oil. Nevertheless, the angular position of peak friction averaged over sliding distances of 0–50 m and 50–100 m remains remarkably stable, appearing consistently near 100° under abrasive wear (Fig. 2(b)) and approximately 90° under base oil (Fig. 2(c)). But under abrasive wear, continuous material removal leads to progressive changes in friction behavior over time, as illustrated by the 100–150 m interval in Fig. 2(b). Therefore, in this study, the angular friction values are based on the average friction within the initial 0–50 m sliding distance, in order to capture the influence of the pristine surface topography on friction.

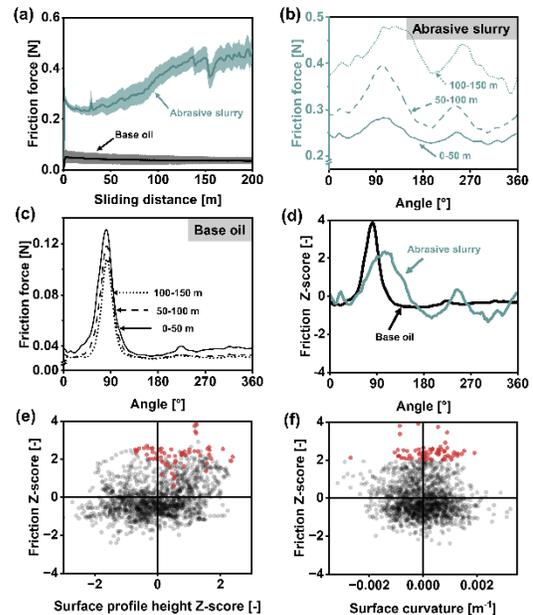

**FIG 2. Standardization of friction force and surface topography.** (a) Friction coefficient versus sliding distance. (b-c) Angle-resolved

friction forces under abrasive wear and base oil conditions at sliding distance intervals of 0–50 m, 50–100 m, and 100–150 m. Peak friction consistently appears near 100° under abrasive wear and approximately 90° under base oil. (d) Dimensionless Z-score enable direct comparison of angle-resolved friction under base oil and abrasive slurry conditions, here shown as frictional data for sliding distances of 0–50 m. (e) Friction Z-score against surface profile height Z-score for all 1560 sections from 13 experiments. The 15° region (5 out of 120 sections) of each disk surface profile with the highest friction is highlighted in red, 74% of these sections have a positive surface profile height Z-score. (f) Friction Z-score plotted against surface curvature. Sections with Z-score greater than two are highlighted in red, 69% of these sections possess a positive curvature.

Then, it is a challenge to identify a statistical method that enables meaningful comparison across friction data exhibiting large quantitative disparities. For example, friction under abrasive slurry and base oil differ significantly when analyzed as a function of time (Fig. 2(a)) or position (Fig. 2(b and c)). To address this, we standardized the friction data by calculating their Z-score. The 132 mm (360°) sliding track in each experiment was divided into 120 sections, and a dimensionless friction Z-score was calculated for each section. The Z-score corresponding to the average friction over 0–50 m, shown in Fig. 2(b and c), are summarized in Fig. 2(d). Friction Z-score and surface profile height Z-score from 13 experiments were calculated (Fig. S3), and data from all sections (1560 in total) were pooled together (Fig. 2(e)). By focusing specifically on the 15° regions (5 out of 120 sections) of each disk's surface profile exhibiting the highest friction forces - marked in red - we observed that 74% of these sections had positive surface profile height Z-score. This suggests that elevated regions in the surface profile are associated with higher friction.

This demonstrates that high peaks on the surface profile are important in the tribological contacts; at the same time, local contact conditions, such as surface curvature, might be critical as well [12]. The surface curvature from the disk surface profile was determined using a low-pass filter with a cutoff length of 7.7 mm, a value deliberately chosen slightly larger than the contact area (7.33 mm), to focus on local contact conditions. The corresponding surface curvature for the surface profile in Fig. 1(b) is shown in Fig. S4. Similarly, we pooled the surface curvature and friction Z-score data for each section from the 13 experiments (Fig. 1(f)). Doing so reveals that 69% of sections with friction Z-score greater than two exhibit positive curvature, indicating a correlation between surface curvature and higher friction. This means that despite our extensive efforts to control and most importantly minimize surface topography to a standard close to that in the semiconductor industry, its influence on friction and wear remains significant.

*Contact author: yulong.li@kit.edu
†Contact author: christian.greiner@kit.edu

## Oscillation's influence on friction

In tribological systems, beyond surface topography, there are further parameters that cannot, realistically, be precisely controlled. For a pin-on-disk configuration, pin inclination can affect friction [13]. Disk alignment [14] is another factor that should be considered. Achieving perfect alignment and true perpendicular normal loading between pin and disk in reality is impossible. Consequently, a minuscule degree of tilt between disk and pin is inevitable, as illustrated in Fig. 3(a). This inherent misalignment in turn leads to what we have termed as oscillations. While making every effort to minimize the amplitude of these oscillations; in our experiments, the oscillation amplitude was still at least 4 μm over a 132 mm sliding track (evaluated with a capacitive sensor in Fig. S5).

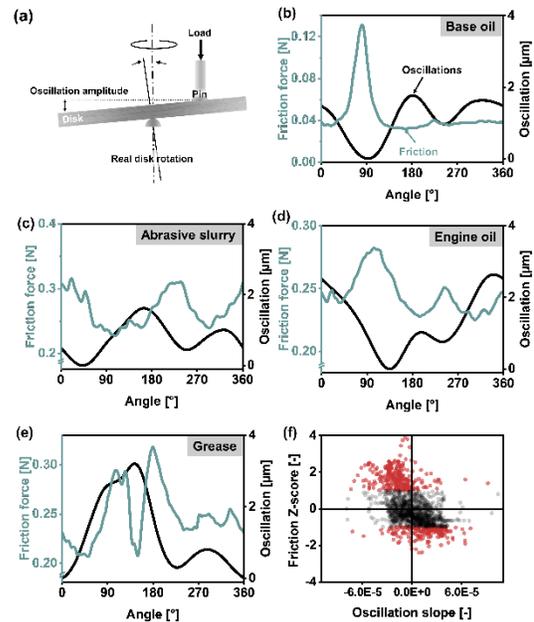

**FIG 3. Comparison of disk oscillation and friction.** (a) Realistic disk rotation scenario, where inevitable tilting during disk mounting results in oscillations of the contacting surfaces. (b-e) Comparing angle-resolved friction with oscillations under different lubrication scenarios - base oil (b), abrasive slurry (c), engine oil (d), and grease (e), using data from the same experiment as in Fig. 1(b-e). (f) Friction Z-score against oscillation slope, using data from the same 13 experiments as in Fig. 1(c and d). Sections where the friction Z-score exceeds one are highlighted in red, and 86% of these regions demonstrate a negative oscillation slope.

The oscillations presented in Fig. 3(b-e) stem from the very same experiment as the topography and friction data shown in Fig. 1(b-e). While Fig. 1(b-e) captures the static surface morphology of the disks, Fig. 3(b-e) reflects their dynamic changes. In comparison to the disk's topography, the correlation between the oscillation profile and the angle-resolved friction forces is not intuitively obvious. To develop such an understanding, we plot the friction Z-score for each disk section from all 13 experiments against the oscillation slope (the slope of each section on the oscillation curve, illustrated in Fig. S4(b); the remaining nine experiments are presented in Fig. S6), as shown in Fig. 3(f), a negative correlation between friction Z-score and the oscillation slope is found. This is particularly true for regions of the disks where friction forces are deviating by more than one standard deviation from the mean value - red data points. When the Z-score exceeds one, 86% of these sections exhibit a negative oscillation slope; conversely, when the Z-score is less than minus one, 80% of regions exhibit a positive oscillation slope. Although the effect of oscillation-induced friction variation has been studied [15], it is remarkable that even under such minimal conditions - where we minimized contact surface oscillations to just 4 µm across a 132 mm sliding track - 86% of high friction sections (with the Z-score exceeding one) from 13 experiments statistically tend to occur when the disk oscillates downwards.

**High Friction Predictor**

Leveraging statistical insights into how initial surface topography and oscillation influence frictional behavior before tribological contact, we sought to establish a model to identify regions on the disk where high friction is likely to occur. With the above results, High Friction Prediction is defined in regions from each disk's 120 sections that simultaneously meet three physical conditions: a negative oscillation slope, a positive surface profile height Z-score, and a positive surface curvature. Even under severely different lubrication scenarios - base oil (Fig. 4(a)), abrasive slurry (Fig. 4(b)), engine oil (Fig. 4(c)), and grease (Fig. 4(d)) - the regions of highest friction consistently tend to overlap with the High Friction Prediction. A clear overlap is observed in the last three cases and close proximity in the case of the base oil. The reliability of this prediction of high friction is further demonstrated across nine repeated experiments in Fig. S7. When plotting the friction Z-score for each disk section from all 13 experiments against the angular distance of each section from the High Friction Prediction (in Fig. 4(e)), we observe that all regions exhibiting high friction (Z-score > 2, red data points) are located within 36° of the High Friction Prediction, with 74% of these concentrated within 12°.

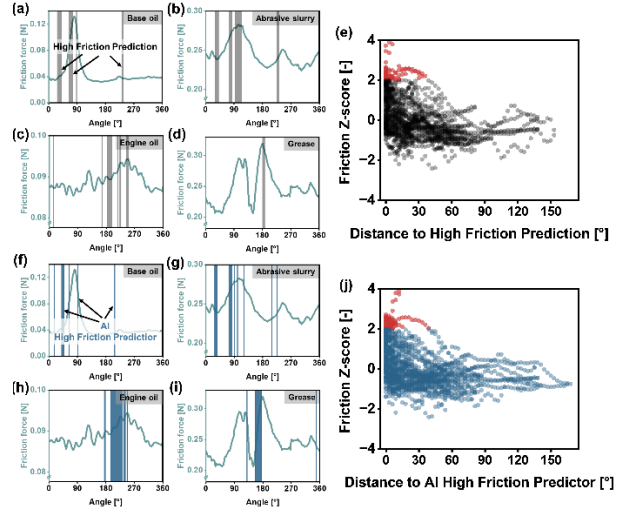

**FIG 4. High Friction Prediction (a-e) and AI High Friction Predictor (f-j).** The four experiments under different lubrication scenarios - base oil (a and f), abrasive slurry (b and g), engine oil (v and h), and grease (d and i) - correspond to the experiments shown in Fig. 1(b-e) and Fig. 2(b-e). (a-d) High friction in a simple model is predicted in sections on the sliding track that simultaneously meet three physical conditions: a negative oscillation slope, a positive surface profile height Z-score, and a positive surface curvature. (f-i) Using a k-nearest neighbors (k-NN) machine learning algorithm for classification, the regions likely to generate high friction are identified with the AI High Friction Predictor. (e) and (j) Friction Z-score for each disk section across all 13 experiments were plotted against their angular distance from the predicted high-friction region (set to zero), using both the High Friction Prediction (e) and the AI High Friction Predictor (j). High friction sections with Z-score > 2 (red) consistently clustered near the predicted regions: 74% within 12° for the High Friction Prediction (e) and 86% within 12° for the AI High Friction Predictor (j).

With the aid of artificial intelligence (AI), we further enhanced the predictive capability. We employed the k-nearest neighbors (kNN) algorithm for classification, using oscillation slope, profile height Z-score, and surface curvature as input features, while the friction Z-score was designated as the target to identify regions where high friction might occur (Friction Z-score > 1). During the machine learning process, all 13 experimental data sets were utilized with 13-fold cross-validation. In this approach, each experiment was sequentially treated as the test set to generate the *AI High Friction Predictor*, while the remaining twelve experiments served as the training set. In a single experiment, comparing the AI predictor (Fig. 4 (f-i)) and the simple model (Fig. 4(a-d)) - or the corresponding experiments shown in Fig. S7 and 8 - reveals similarities; but it is challenging to evaluate which performs better. Therefore, in Fig. 4(j), the friction Z-score for each disk section across all 13 experiments was plotted against the angular distance of each section from the *AI High Friction Predictor*, we observed that 86% of the high friction regions (Z-score > 2, red data points) were concentrated within 12°. This represents a 12% improvement over our classical model (the High Friction Prediction in Fig. 4(e)). Across all 13 experiments, the AI High Friction Predictor accurately located the highest friction points in eleven experiments.

*Contact author: yulong.li@kit.edu
†Contact author: christian.greiner@kit.edu

### III. DISCUSSION

Measuring surface topography, particularly roughness, and considering its effect on friction and wear is a long standing focus in tribology. Therefore numerous successful models have been developed and adopted to predict tribological behavior [16,17]. Yet, even when all (seemingly) relevant parameters are specified, pinpointing where the highest friction will occur on stringently controlled surfaces remains very challenging. In turn, this is exactly what is needed to know where a tribological surface should be prepared differently, or where the system most likely will fail due to excessive wear. Our experimental findings highlight the complexity behind this goal and why it has not yet been achieved. We have identified two essential surface features - the profile height and surface curvature - that are necessary for predicting high friction. However, the relationship between these parameters and the frictional forces remains somewhat chaotic, as neither factor alone robustly predicts high friction (see Fig. 1(b-e)).

This complexity arises largely because misalignment-induced oscillations, even when minimized, continuously influence friction behavior. The reproducibility challenge in friction can arise from surface topography and oscillations and their complex interactions which may ultimately give rise to system-level chaotic behavior. This chaotic nature is evidenced by positive Lyapunov exponents [18] calculated for the friction data shown in Fig. 1(b-e). The Lyapunov exponent, while it has to be used with caution, can be applied to hint towards the predictability of a system [18]. For the time-series friction data in Fig. 1(b-e), the largest Lyapunov exponents are 0.38, 0.16, 0.18, and 1.45, respectively. These positive values indicate a significant degree of chaos within the tribological system [19]. This is not surprising, as even a simple dry-friction oscillator model is known to exhibit chaotic behavior [20]. Future research might chart a promising path toward models akin to weather forecasting, where integrating diverse scales and factors enables predictive insight into tribological behavior in technologically relevant systems.

Our data demonstrate that three physical conditions - a negative oscillation slope, a positive surface profile height Z-score, and a positive surface curvature - contribute to the occurrence of high friction. While surface topography [12] and oscillations [21] are not new to tribological studies, elucidating the mechanisms underlying the physical conditions we identified remains a challenge that calls for collective efforts from the community. Nevertheless, it is possible to develop meaningful heuristics (High Friction Prediction in Fig. 4(a-d) and Fig. S7) to, for example, predict where the highest friction forces will be present on a given surface profile. By leveraging AI, these predictions can be further refined (AI High Friction Predictor in Fig. 4(f-i) and Fig. S8). Our findings have been achieved for a wide range of tribological environments, using interfacial media from unadditived base oils to aerospace greases, fully formulated engine oils, and abrasive slurries. While these conditions might seem to amplify the already chaotic nature of tribological systems, it is surprising to find such clear commonalities between them. The AI-based prediction of high friction regions works surprisingly well despite the generally chaotic nature of tribology. For example and very strikingly, peak friction under grease lubrication (in Fig. 4(i)) can be accurately inferred using data from experiments with three entirely different lubricants.

*Contact author: yulong.li@kit.edu
†Contact author: christian.greiner@kit.edu

In processing the surface profile and oscillation data, we deliberately suppressed high-frequency small-scale roughness (e.g., the surface profiles analyzed had a spatial resolution of 1.1 mm). Unlike traditional models that focus on small-scale roughness [16,17], our statistical results demonstrate that mid-scale topography, together with misalignment-induced oscillations, can successfully be used to predict where friction will be highest. This complements recent global collaborative findings underscoring the critical role of scale in roughness analysis [22] and extends that principle directly to predictive capabilities. Academically, our work offers new perspectives on how mid-scale topography is coupled with friction. Practically, high friction and wear can be predicted with over 80% accuracy from surface topography measurements at only 1.1 mm resolution, without the need for elaborate experimental techniques - making this approach readily applicable to a strategic surface design and proactive maintenance strategies in engineering practice.

### ACKNOWLEDGMENTS

We express our gratitude to J. Schneider and N. Garabedian for lab help and discussions. CG would like to acknowledge financial support from European Research Council (ERC) grant (771237) and German Research Foundation grant (GR 4174/12).

### APPENDIX

Materials, methods and evaluation details are in **Supplemental Material**.

[1] K. Holmberg and A. Erdemir, Influence of tribology on global energy consumption, costs and emissions, Friction **5**, 263 (2017).
[2] R. W. Carpick, A. Jackson, W. G. Sawyer, N. Argibay, P. Lee, A. Pachon, and R. M. Gresham, The tribology opportunities study: can tribology save a quad?, Tribology & Lubrication Technology **72**, 44 (2016).
[3] M. Watson et al., An analysis of the quality of experimental design and reliability of results in tribology research, Wear **426–427**, 1712 (2019).
[4] C. E. Morstein, A. Klemenz, M. Dienwiebel, and M. Moseler, Humidity-dependent lubrication of highly loaded contacts by graphite and a structural transition to turbostratic carbon, Nat Commun **13**, 5958 (2022).
[5] C. Haug, F. Ruebeling, A. Kashiwar, P. Gumbsch, C. Kübel, and C. Greiner, Early deformation mechanisms in the shear affected region underneath a copper sliding contact, Nat Commun **11**, 839 (2020).
[6] V. Slesarenko and L. Pastewka, The bumpy road to friction control, Science **383**, 150 (2024).
[7] H. Czichos, S. Becker, and J. Lexow, Multilaboratory tribotesting: Results from the Versailles Advanced Materials and Standards programme on wear test methods, Wear **114**, 109 (1987).
[8] T. D. B. Jacobs, L. Pastewka, and Guest Editors, Surface topography as a material parameter, MRS Bulletin **47**, 1205 (2022).
[9] A. R. Hinkle, W. G. Nöhring, R. Leute, T. Junge, and L. Pastewka, The emergence of small-scale self-affine surface roughness from deformation, Sci. Adv. **6**, eaax0847 (2020).
[10] R. Aghababaei, E. E. Brodsky, J.-F. Molinari, and S. Chandrasekar, How roughness emerges on natural and engineered surfaces, MRS Bulletin **47**, 1229 (2022).
[11] S. Iida, T. Nagai, and T. Uchiyama, Standard wafer with programed defects to evaluate the pattern inspection tools for 300-mm wafer fabrication for 7-nm node and beyond, J. Micro/Nanolith. MEMS MOEMS **18**, 1 (2019).
[12] L. Pastewka and M. O. Robbins, Contact between rough surfaces and a criterion for macroscopic adhesion, Proc. Natl. Acad. Sci. U.S.A. **111**, 3298 (2014).
[13] H. Yue, J. Schneider, B. Frohnapfel, and P. Gumbsch, The influence of pin inclination on frictional behaviour in pin-on-disc sliding and its implications for test reliability, Tribology International **200**, 110083 (2024).
[14] I. Garcia-Prieto, M. D. Faulkner, and J. R. Alcock, The influence of specimen misalignment on wear in conforming pin on disk tests, Wear **257**, 157 (2004).
[15] M. A. Chowdhury and M. Helali, The effect of amplitude of vibration on the coefficient of friction for different materials, Tribology International **41**, 307 (2008).
[16] M. Marian, M. Bartz, S. Wartzack, and A. Rosenkranz, Non-Dimensional Groups, Film Thickness Equations and Correction Factors for Elastohydrodynamic Lubrication: A Review, Lubricants **8**, 95 (2020).
[17] Contact of nominally flat surfaces, Proc. R. Soc. Lond. A **295**, 300 (1966).
[18] M. T. Rosenstein, J. J. Collins, and C. J. De Luca, A practical method for calculating largest Lyapunov exponents from small data sets, Physica D: Nonlinear Phenomena **65**, 117 (1993).
[19] C. Ding, H. Zhu, G. Sun, Y. Zhou, and X. Zuo, Chaotic characteristics and attractor evolution of friction noise during friction process, Friction **6**, 47 (2018).
[20] G. Licskó and G. Csernák, On the chaotic behaviour of a simple dry-friction oscillator, Mathematics and Computers in Simulation **95**, 55 (2014).
[21] T. Dimond, A. Younan, and P. Allaire, A Review of Tilting Pad Bearing Theory, International Journal of Rotating Machinery **2011**, 1 (2011).
[22] A. Pradhan et al., The Surface-Topography Challenge: A Multi-Laboratory Benchmark Study to Advance the Characterization of Topography, Tribol Lett **73**, 110 (2025).
[23] C. Greiner, T. Merz, D. Braun, A. Codrignani, and F. Magagnato, Optimum dimple diameter for friction reduction with laser surface texturing: the effect of velocity gradient, Surf. Topogr.: Metrol. Prop. 3, 044001 (2015).
[24] L. Pastewka and M. O. Robbins, Contact between rough surfaces and a criterion for macroscopic adhesion, Proc. Natl. Acad. Sci. U.S.A. **111**, 3298 (2014).
[25] M. D. Weir, J. Hass, and G. B. Thomas, Thomas' Calculus: Early Transcendentals, Thirteenth edition (Pearson, Boston, 2014).
[26] N. J. Salkind and K. Rasmussen, editors , Encyclopedia of Measurement and Statistics (SAGE Publications, Thousand Oaks, Calif, 2007).
[27] E. N. Lorenz, Deterministic Nonperiodic Flow, J. Atmos. Sci. **20**, 130 (1963).
[28] F. Takens, Detecting Strange Attractors in Turbulence, in Dynamical Systems and Turbulence, Warwick 1980, edited by D. Rand and L.-S. Young, Vol. 898 (Springer Berlin Heidelberg, Berlin, Heidelberg, 1981), pp. 366–381.
[29] M. T. Rosenstein, J. J. Collins, and C. J. De Luca, A practical method for calculating largest Lyapunov exponents from small data sets, Physica D: Nonlinear Phenomena 65, 117 (1993).
[30] Z. Zhang, Introduction to machine learning: k-nearest neighbors, Ann. Transl. Med. 4, 218 (2016).
[31] Y. Li, N. Garabedian, J. Schneider, and C. Greiner, Waviness Affects Friction and Abrasive Wear, Tribol Lett **71**, 64 (2023).

*Contact author: yulong.li@kit.edu
†Contact author: christian.greiner@kit.edu